\documentclass[%
preprint,
frontmatterverbose, 
nofootinbib,
nobibnotes,
amsmath,amssymb,
aps,
pra,
floatfix,
superscriptaddress,
longbibliography
]{revtex4-1}
\usepackage[utf8]{inputenc}
\usepackage[T1]{fontenc} 
\usepackage{amsmath,amssymb}
\usepackage{graphicx}
\usepackage{bm}
\usepackage{miller}
\usepackage{textcomp}
\usepackage{siunitx} 
\usepackage{comment}

\newcommand{\degree}{$^{\circ}$}

\usepackage[unicode=true, 
 bookmarks=true,bookmarksnumbered=false,bookmarksopen=false,
 breaklinks=true,pdfborder={0 0 1},backref=false,colorlinks=true]
 {hyperref}
\hypersetup{pdftitle={Kikuchi Pattern Spectrum},
 pdfauthor={Aimo Winkelmann}}

\begin{document}

\title{Constraints on the Effective Electron Energy Spectrum in Backscatter Kikuchi Diffraction}

\author{Aimo Winkelmann}
\email{a.winkelmann@lzh.de}
\affiliation{Laser Zentrum Hannover e.V., Hollerithallee 8, 30419 Hannover, Germany }

\author{T. Ben Britton}
\email{b.britton@imperial.ac.uk}
\affiliation{Department of Materials, Imperial College London, London, UK}

\author{Gert Nolze}
\affiliation{Federal Institute for Materials, Research and Testing (BAM), Unter den Eichen 87, 12205 Berlin, Germany}
\email{gert.nolze@bam.de}

\date{\today}

\begin{abstract}
    Electron Backscatter Diffraction (EBSD) is a technique to obtain microcrystallographic information from materials by collecting large-angle Kikuchi patterns in the scanning electron microscope (SEM).
    An important fundamental question concerns the scattering-angle dependent electron energy distribution which is relevant for the formation of the Kikuchi diffraction patterns. 
    Here we review the existing experimental data and explore the effective energy spectrum that is operative in the generation of backscatter Kikuchi patterns from silicon. We use a full pattern comparison of experimental data with dynamical electron diffraction simulations.
    Our energy-dependent cross-correlation based pattern matching approach establishes improved constraints on the effective Kikuchi pattern energy spectrum which is relevant for high-resolution EBSD pattern simulations and their applications. 
\end{abstract}


\maketitle

\section{Introduction}

Electron backscatter diffraction (EBSD) is a technique which is used to reveal the microstructure of crystalline materials, including metals, ceramics, functional materials, and minerals in the scanning electron microscope (SEM) \cite{schwartzEBSD2}. 
Recently, there has been a growth in the number of studies which use high quality pattern simulations \cite{winkelmann2007um,maurice2011um,callahan2013mm,liu2016jm,cheng2018njp} to expand the application areas of EBSD and to render new insight into the microstructure of materials (e.g.  \cite{naresh2017srep,singh2018srep,wilkinson2019um}).
In order to use these simulations with confidence, we need to ensure that the relevant electron scattering mechanisms and diffraction physics are correctly included in our theoretical models of EBSD pattern formation.

\begin{figure}[bt!] %
	\centering
	  	\includegraphics[width=0.8\textwidth,trim={0cm 0 0cm 0cm},clip]{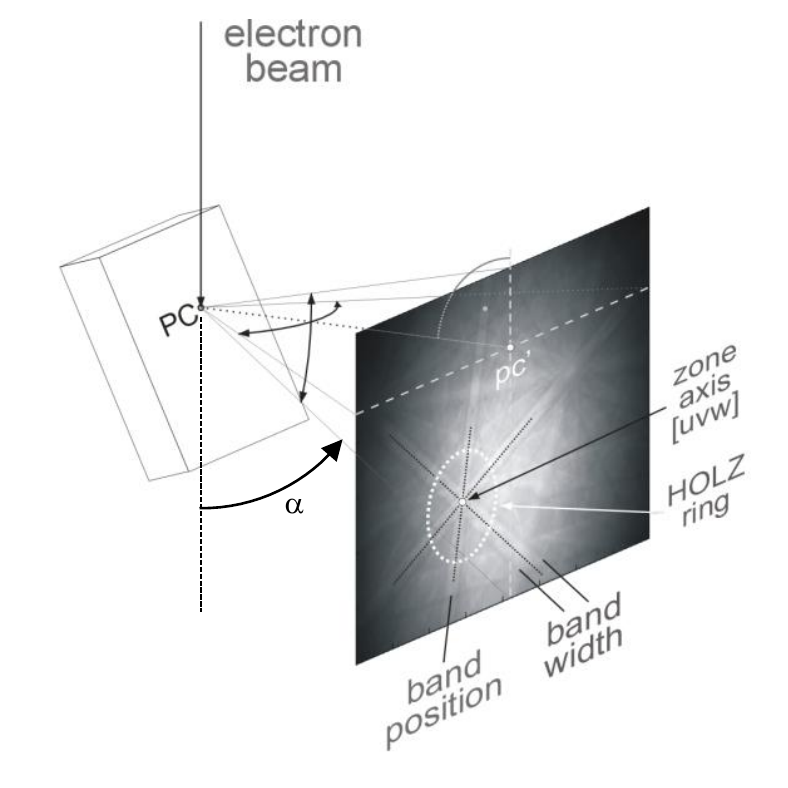}
	\caption{EBSD scattering geometry and raw diffraction pattern with key features.  For a specific point on the phosphor screen, the angle $\alpha$ indicates the scattering angle relative to the primary beam, with $30^\circ \lessapprox \alpha \lessapprox 130^\circ$ for typical EBSD patterns.}
	\label{fig:s_ebsd}
\end{figure}

An important fundamental question concerns the scattering-angle dependent electron energy distribution which is relevant for EBSD patterns, where the effective scattering angles can change by values in the order of 90\,\degree{} within a single diffraction pattern (see Fig.\,\ref{fig:s_ebsd}).
Because it is known that the backscattered electron energy spectrum can change considerably with scattering angle \cite{wellssem,reimersem,goldsteinSEM4}, it is crucial to have an explicit understanding about the ensuing changes in the energy distribution of those particular electrons that convey the crystallographic information via the observed diffraction features.  
Some constraints on the energy spectrum which is relevant for EBSD Kikuchi patterns have been previously established by spectroscopic measurements \cite{deal2008um,winkelmann2010njp,vespucci2015prb} and by comparison of experimental interference features to dynamical electron diffraction simulations \cite{winkelmann2009ebsd2}.
With continuing progress in quantitative Kikuchi pattern simulations for EBSD applications \cite{winkelmann2016iop}, in this paper we revisit the problem of the energy dependence in EBSD patterns. 
Using quantitative image comparisons, we will investigate whether the currently available experimental data on the EBSD energy spectrum is consistent with experimentally measured wide-angle EBSD diffraction patterns which we compare to energy-dependent Kikuchi diffraction pattern simulations.
In contrast to the recent study of \citeauthor{ram2018prb} \cite{ram2018prb}, we find that Kikuchi patterns from silicon are consistent with mean energies which are approximately 1 to 1.5\,keV below the primary beam energy, compared to a corresponding range between 2 and 5\,keV predicted in \cite{ram2018prb}. 
Energy differences of this size would have considerable impact especially on high resolution EBSD methods for strain determination, and therefore it is important to resolve the apparent inconsistency.
We assign the central source of the discrepancy concerning the effective Kikuchi pattern spectrum to the use of the continuous slowing down approximation (CSDA) in the Monte Carlo simulations of the electron energy spectrum as presented in \cite{callahan2013mm,ram2018prb,singh2018srep}.

\section{Essential Background}

\subsection{Experimental Geometry for Kikuchi Pattern Measurements}

In order to explain the experimental scattering geometry, a typical set up for EBSD analysis is shown schematically in Fig.\,\ref{fig:s_ebsd}. 
A crystalline sample is tilted to a high angle (often near 70\si{\degree}) in the SEM to increase the yield of electrons that are backscattered from the sample and undergo diffraction effects. The backscattered electrons (BSE) are captured using a flat screen, with scattering angles $\alpha$ ranging typically from 30...40\degree{} at the bottom of the screen to 110...130\degree{} at the top.
The diffraction patterns which are observed in EBSD are Kikuchi patterns \cite{kikuchi1928praj,alam1954prsa} formed by incoherent sources inside a crystal.
The formation of these incoherent sources relative to the incident wave is related to the localized recoil of single atoms in the backscattering process of an electron, as has been shown by spectroscopic, element-resolved, diffraction measurements \cite{winkelmann2011prl}.

\begin{figure}[bt!]  %
	\hspace{0.78cm} 
	\includegraphics[width=0.98\textwidth]{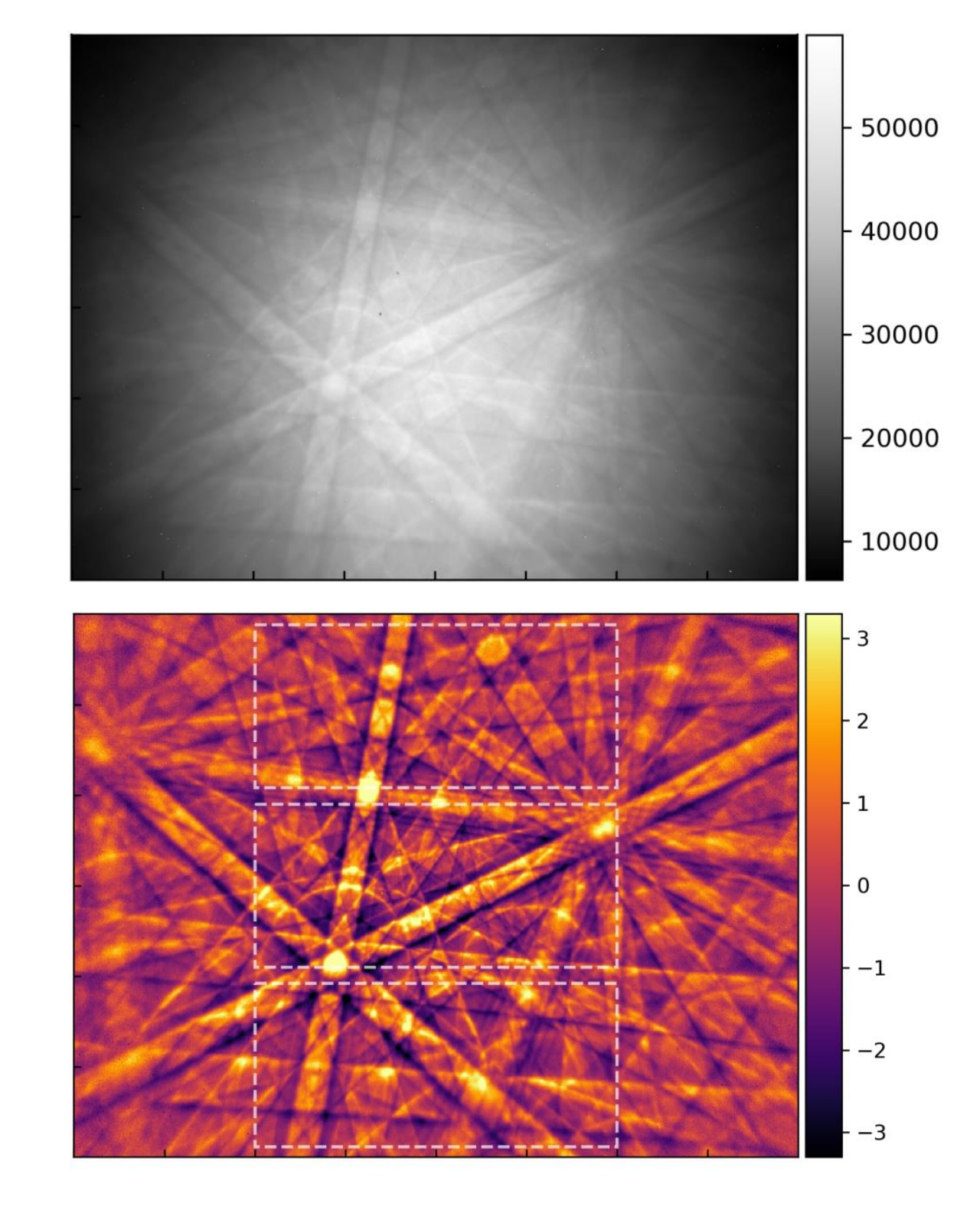}
	\caption{Upper Panel: experimental raw EBSD pattern from silicon, $E_0=15$\,keV. 
	Lower Panel: background-processed silicon Kikuchi pattern. The dashed rectangles mark the regions of interest corresponding to different ranges of scattering angles. Projection center: \mbox{(0.5024,  0.1555,  0.7751) \cite{britton2016mc}}, capture angles: horizontally: 81\degree, vertically: 59\degree}
	\label{fig:sikiku}
\end{figure}

With respect to the experimental data acquisition, we show in the top part of Fig.\,\ref{fig:sikiku} an experimental, raw EBSD pattern before the subsequent image processing that uncovers the inherent Kikuchi pattern shown below. 
The processed pattern has been obtained by removing the slowly varying diffuse background from the raw data and normalizing the pattern intensity to a mean $\mu=0.0$ and standard deviation $\sigma=1.0$, as indicated by the intensity scale. The background removal procedure involves an initial division of the raw pattern by a static background obtained from the aluminum sample holder, followed by the division by a dynamic (per pattern) background obtained by low-pass FFT filtering of the pattern obtained after the first step. 
For the majority of EBSD analysis methods, the processed pattern and not the raw pattern is used to extract the actual crystallographic information.
Compared to the processed pattern, the raw pattern can be strongly influenced by changes in the overall backscattering coefficient from the sample due to local variations in density, surface topography, shadowing, and incident beam diffraction \cite{winkelmann2017jm}, among other effects, which can interfere with the determination of crystal phase and orientation from the EBSD Kikuchi pattern. 
If a raw pattern simulation is carried out, the simulated diffuse background can be removed in the same way as shown for the experiment in Fig.\,\ref{fig:sikiku} in order to quantitatively compare simulation and experiment. 
This implies that simple models for the diffuse background might be sufficient if it is subsequently removed anyway.

The geometry of the Kikuchi patterns is governed by the gnomonic projection of the conditions for Bragg reflection of waves originating from a point source inside a crystal \cite{Kossel1936a}. 
The width of the Kikuchi bands seen in  Fig.\,\ref{fig:sikiku} is related to the Bragg angle determined by the respective lattice spacing $d_{hkl}$ and the wavelength $\lambda$ of the scattered electrons.
Basic crystallographic diffraction theory \cite{degraef2007structure} predicts that lower electron energies lead to larger Bragg angles and thus to an increase of the width of the Kikuchi bands and a change in the location of the intersections of the band edges. Lower electron energies will also increase the diameter of the higher order Laue zone (HOLZ) rings \cite{michael2000um} seen around the zone axes at the intersections of strong Kikuchi bands.
For a finite spectrum of BSE energies, we will thus have a superposition of Kikuchi pattern features at different energies, which will lead to a corresponding broadening of features in the detected EBSP.
Moreover, we can also expect that the energy spectrum can vary with position in the Kikuchi pattern, as the BSE spectrum depends on the scattering angle.
Therefore, we have indicated three rectangular regions of interest (ROI) in Fig.\,\ref{fig:sikiku}, where the lower ROI corresponds to the smallest scattering angles, while the upper ROI is related to significantly larger scattering angles (compare the experimental setup shown in Fig.\ref{fig:s_ebsd}).
One of the main subjects of this paper will be to determine which size of variations in the electron energy spectrum are compatible with the observed diffraction features in the different ROIs.

\subsection{Energy dependence of EBSD patterns}

Theoretical models of EBSD have to consider a spectrum of backscattered electron energies in the treatment of the dynamical diffraction effects, as discussed in \cite{winkelmann2009ebsd2,eades2009ebsd2}, for example.

The properties of the BSE spectrum in the SEM have been studied in the past \cite{wellssem,wolf1973jvst,darlington1975jpd,niedrig1978sca,shorter1980thesis,reimersem,goldsteinSEM4}.
Without going into too much theory, we can expect that the intensity of the BSEs depends on their energy and scattering angle, and also the relative strength of the diffraction effects will be a function of these parameters.
This means that the relative modulation of interference features can vary with the electron energy, and we could hypothesize, for example, that the electrons with increasing energy loss will show less and less diffraction information because inelastic scattering tends to destroy coherence of the multiply scattered electron waves.

This hypothesis is consistent with the available experimental investigations in which the spectral dependence of SEM diffraction effects has been directly studied by measuring the angle-resolved and energy-dependent BSE intensity with high spectral resolution.

\citeauthor{berger2002scanning} \cite{berger2002scanning} have investigated diffraction effects of the incident beam on the BSE spectrum of Si at 20\,keV and demonstrated that the corresponding changes in the BSE spectrum occur within approx. 1\,keV of the primary beam energy.

\citeauthor{deal2008um} \cite{deal2008um} used an electrostatic high-pass imaging filter to measure energy-dependent Kikuchi patterns from Si, Fe, and Ir, at a primary beam energy of 15\,keV, see also \cite{bhattacharyya2009scanning}.
In their analysis, \citeauthor{deal2008um} found that the contribution of electrons to the Kikuchi patterns decreases with energy loss, i.e.\@ electrons with large energy losses contribute mainly to the diffuse background intensity in the raw pattern.
\citeauthor{deal2008um} conclude that the major contributors to the EBSD patterns are electrons with approximately 97\% of the incident beam energy \cite{deal2008um} (i.e.\@ a loss in the order of 0.5\,keV for a 15\,keV primary beam energy).

\citeauthor{winkelmann2010njp} \@ \cite{winkelmann2010njp} performed angle-resolved electron energy loss measurements for 30\,keV electrons backscattered from silicon, with an energy resolution below 1\,eV, which allowed to distinguish the specific contribution of plasmon losses to the diffraction pattern. 
The analysis of the diffraction modulation as a function of energy showed that high Kikuchi band contrast is associated with energy losses approximately in the range below 1\,keV for scattering angles of 135\degree{} and an angle of incidence of 75\degree{}.
The maximum in band contrast was shown to depend on the relative path lengths in the sample. When the inelastic scattering happens predominantly on the incident part of the path before the backscattering event, the band contrast can still be high even for higher numbers of plasmon losses. 
It was also shown that the Kikuchi profile becomes blurred by reduced dynamical scattering at low energy losses when the geometrical conditions are such that the effective sample thickness for the backscattered electrons is very low ($<5$nm).

Based on these spectroscopic investigations, the Kikuchi pattern simulation model discussed in \cite{winkelmann2007um,winkelmann2009ebsd2, winkelmann2016iop} makes the approximation that we can effectively divide the backscattered electrons in two groups: those that contribute mainly to the Kikuchi diffraction pattern, and those electrons that mainly contribute to the diffuse background.
In general, these two groups can have different energy spectra, because of the quantitatively different buildup of the respective multiple scattering processes in both cases.
Compared to a possibly very broad (multiple keV) BSE spectrum in the diffuse background, the effective Kikuchi pattern spectrum is assumed to be narrow and peaked in the vicinity below the primary beam energy $E_0$ (depending on the material and scattering geometry up to about 1\,keV below $E_0$)  \cite{winkelmann2010njp}, with an effective width in the order of $\lessapprox1$\,keV. 
Under these conditions, it is assumed that the remaining effects of energy spread can be described by an empirical instrumental broadening of the simulated diffraction patterns for a single mean energy, or by averaging a number of diffraction patterns within a small range near the mean energy. 

Using energy-resolved pattern simulations within this approximation, it was shown in \cite{winkelmann2009ebsd2} that EBSD patterns observed from a GaN sample at 20\,keV primary beam energy can be described by averaging over an effectively constant electron spectrum from 19.5 to 20\,keV, including an additional instrumental angular broadening. By comparison to experimental features of crossing lines near a HOLZ ring, it was also shown that for a pattern simulated at 18.5\,keV, we can already observe a clear deviation from the experiment, excluding a larger energy range for the effective Kikuchi pattern spectrum in the described experiment.     
 
If the energy spread of the backscattered electrons as a function of position across the detector would be the dominating mechanism that causes changes of diffraction features, we could also analyse the width of experimental Kikuchi bands directly to obtain information on the corresponding electron energy spectrum \cite{alam1954prsa,shorter1981jpd,dingley2004jm}.
Using this indirect, non-spectroscopic, approach, \citeauthor{ram2018prb} \cite{ram2018prb} analyzed the apparent widening of a single selected Kikuchi band from a silicon sample.
Using  Monte Carlo (MC) simulations of electron scattering, \citeauthor{ram2018prb} suggested that the mean energy of the electrons which strike the detector and form the diffraction pattern depends strongly on the scattering angle and therefore there should be a large variation in energy across a 2D EBSD detector (as the EBSD detector typically subtends a large capture angle).
In \cite{ram2018prb}, the electrons in these angle-dependent BSE spectra are simulated according to the continuous slowing down approximation (CSDA) \cite{joy1995mc} and contribute to the Kikuchi pattern according to their relative spectral intensity \cite{callahan2013mm}, i.e. the CSDA-MC diffuse background for a specific energy is multiplied by the diffraction modulation from a dynamical electron diffraction simulation.

The predictions of the CSDA-MC simulations in \cite{ram2018prb} are validated by comparison of the apparent width of one Kikuchi band extracted from the silicon pattern using a Fourier filtering method \cite{ram2014jac}.
For an EBSD pattern from silicon at a primary incoming beam energy of 15\,keV, the results in \cite{ram2018prb} seem to suggest that the mean energies which contribute to the Kikuchi pattern are in a range from 13\,keV at the bottom of the pattern to below 11\,keV at the top of the pattern.
Furthermore, the BSE spectra presented in \cite{ram2018prb} for different positions on the phosphor screen show an increase in their spread, as we characterize here by the full width at half maximum (FWHM) from the peak energy. In the data shown in \cite{ram2018prb}, the FWHM increases from about $2.5$\,keV to $>6$\,keV from bottom to top of the pattern.
The geometrical conditions for the scattering angles at the top of the EBSD pattern, with the largest losses and largest broadening of the spectrum, approximately correspond to the geometry with 75\degree{} angle of incidence studied spectroscopically in \cite{winkelmann2010njp} as discussed above.

\subsection{Approach used in the current study}

In the current paper, we use full pattern dynamical electron diffraction simulations to explore the possible impact of the BSE energy spectrum on the appearance of EBSD patterns from silicon. 
Compared to an averaged band width, the crossing points of features in dynamical pattern simulations are very sensitive to energy, which can be used, for example, to calibrate the beam voltage in quantitative convergent beam electron diffraction (CBED) in the transmission electron microscope \cite{carterwilliams2016companion}. An energy sensitivity of the same kind  allow us to directly judge the fit of simulated Kikuchi patterns to the experiment visually, in addition to a quantitative numerical image similarity measurement via the normalized cross-correlation coefficient (see below).
Observation of features of crossing lines in the Kikuchi patterns can also be seen as a consistent band width measurement of multiple bands, because the visible linear features correspond to Kikuchi band edges. It is the change in the Kikuchi band widths that determines the relative appearance of the crossing line features. 
As discussed above, a previous analysis of high-resolution features in a GaN EBSD pattern has been carried out in \cite{winkelmann2009ebsd2} for a limited range of scattering angles. In the current paper, the extension of this previous approach to several ROIs simultaneously will allow us to precisely estimate the energy-dependent effects in full Kikuchi patterns.

\section{Experimental and Theoretical Details}

\subsection{EBSD pattern measurement}

The silicon EBSD data was measured from a sample of a commercial Si(001) wafer at 15kV acceleration voltage.
Maps of $50\times 37$ patterns with a resolution $800\times 600$ pixels using an e$^-$Flash$^\textrm{HR}$ EBSD detector (Bruker Nano) were acquired at 6000x magnification using a FE-SEM LEO\,1530VP at 10\,nA probe current and 400ms exposure time in the high vacuum mode.

After the measurement, 10 patterns near a selected position were averaged to optimize the signal-to-noise ratio.
Background processing of the raw EBSD patterns was done via static and dynamic background division as discussed above to result in an approximately constant average intensity. For quantitative image visualization, we normalize the Kikuchi patterns to the mean of $\mu=0.0$ and standard deviation of \mbox{$\sigma=1.0$}. The same normalization is applied to the simulated patterns. 
No further changes in brightness or contrast were applied.

The measurements of the BaFe$_2$As$_2$ patterns \cite{pukenas2018micron} were done on a Zeiss Ultra 55 SEM with a Nordlys HKL EBSD system, and a low-temperature stage at 12\,K. The acceleration voltage was 20\,keV, with a probe current of 10..11nA (SEM aperture size 120$\mu$m). 
The sample was tilted at 70\degree, the detector distance was 16.5\,mm. Pattern averaging was done for 10 patterns at 43\,ms capture time.

\subsection{Energy-Dependent EBSD Pattern Simulation}

Simulation of the dynamical master data is performed for a specified electron energy spread and the assumed crystal structure according to the Bloch wave approach described in \cite{winkelmann2007um}.
We do not model the diffuse background like in \cite{callahan2013mm}, as we select to background-process our raw experimental patterns. 
Our model gives the relative variation of the diffracted intensity with respect to completely incoherent backscattering from atomic scatterers without any diffraction effects.
We have calculated the master data for silicon in the range of 11.5 to 15.5\,keV in steps of 100eV and then averaged the master data according to a Gaussian distribution with given full width at half maximum (FWHM). For the analysis shown below, we have used a simulated diffraction spectrum with a FWHM of 500\,eV.
In addition, we have applied to the master data an instrumental broadening through convolution of the simulated pattern with a Gaussian filter of approx.\@ 1.5\,mrad resolution \cite{winkelmann2009ebsd2}. 
This resulted in 6 sets of master data for silicon from 12.5 to 15.0 keV in steps of 500\,eV. 
In the dynamical calculation we included a total of 2222 reflectors with minimum lattice spacing $d_{hkl}>0.35$\AA. The Debye-Waller-Factor for Si was taken as $B=0.8$\,\AA$^{-2}$.

For the simulations of the BaFe$_2$As$_2$ patterns, we have used 2936 reflectors with a minimum lattice spacing $d_{hkl}>0.35$\AA, and a Debye-Waller-Factor of $B=0.3$\,\AA$^{-2}$ to account for the experimental temperature of 12\,K. We calculated 8 sets of master data, at energies from 18.0\,keV to 21.5\,keV in steps of 0.5\,keV, without additional energy broadening or angular smoothing.

Concerning the depth distribution that is necessary for the dynamical simulations, we use analytical distributions which take into account the qualitative features of the Kikuchi source distribution \cite{winkelmann2010jm}. Most importantly, the mean depths from which the Kikuchi pattern electrons are emitted, have to be in the order of the inelastic mean free path (IMFP), as inelastic scattering on the way out of the crystal will destroy coherence. 
For the depth dependent source strength in silicon, we use a Poisson-type profile $ \propto t/t_\textrm{KIK}\cdot\exp(t/t_\textrm{KIK})$ with the mean depth of excitation  $t_\textrm{KIK} = 13$nm and an IMFP $\lambda_\textrm{IMFP} = 15$nm.
For BaFe$_2$As$_2$ we have used $t_\textrm{KIK}=6$nm  and $\lambda_\textrm{IMFP} = 8$nm.

Quantitative image comparison is performed by calculating the normalized cross-correlation coefficient (NCC) $r$ ($0<|r|<1$) \cite{gonzalez2007DIP3} between two Kikuchi patterns.

The projection center was determined from the best full pattern fit of orientation and projection center for all 6 assumed mean energies. Based on the high pattern resolution of $800\times 600$ pixels, the accuracy of the PC value determined in this way is estimated to be in the order of less than 0.1\% and is not expected to influence the final result.
In the subsequent fit for the respective regions of interest, the same, fixed PC was used for all energies and optimization of the NCC was performed through small variations in the orientation for each energy, to find the best match between experiment and simulation.

\section{Results}

\begin{figure}[tb!]  %
    \includegraphics[width=0.9\textwidth]{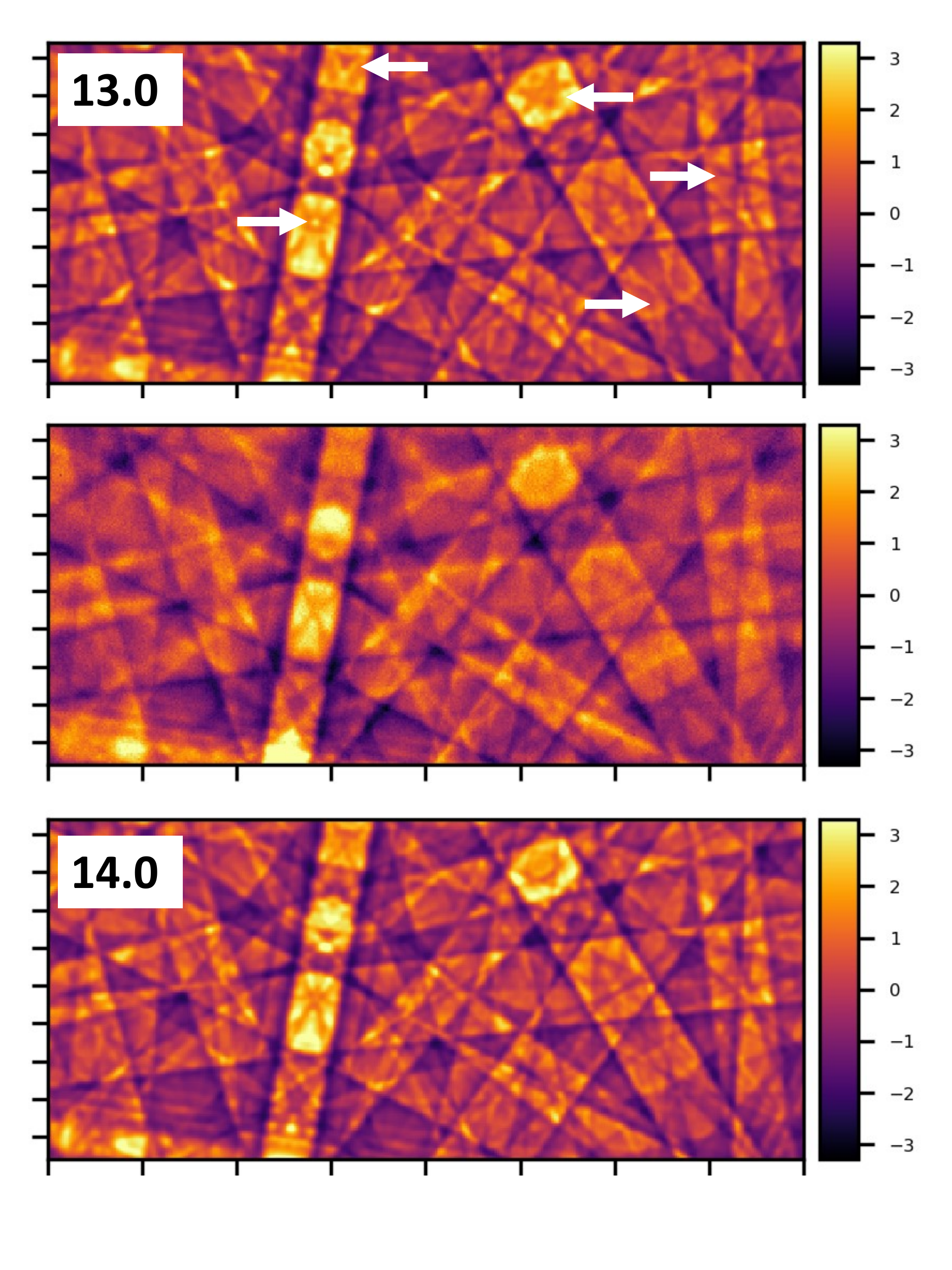}
	\caption{Analysis of the experimental upper ROI. Top: simulation for 13.0\,keV ($r=0.839$) with indicated features for comparison to the experiment. Center: Experimental  ROI. Bottom: best fit ROI simulation for 14.0\,keV central energy ($r=0.850$), features indicated in the upper panel fit better at 14\,keV.}
	\label{fig:roi_top}
\end{figure}

\begin{figure}[tb!]  %
    \includegraphics[width=0.9\textwidth]{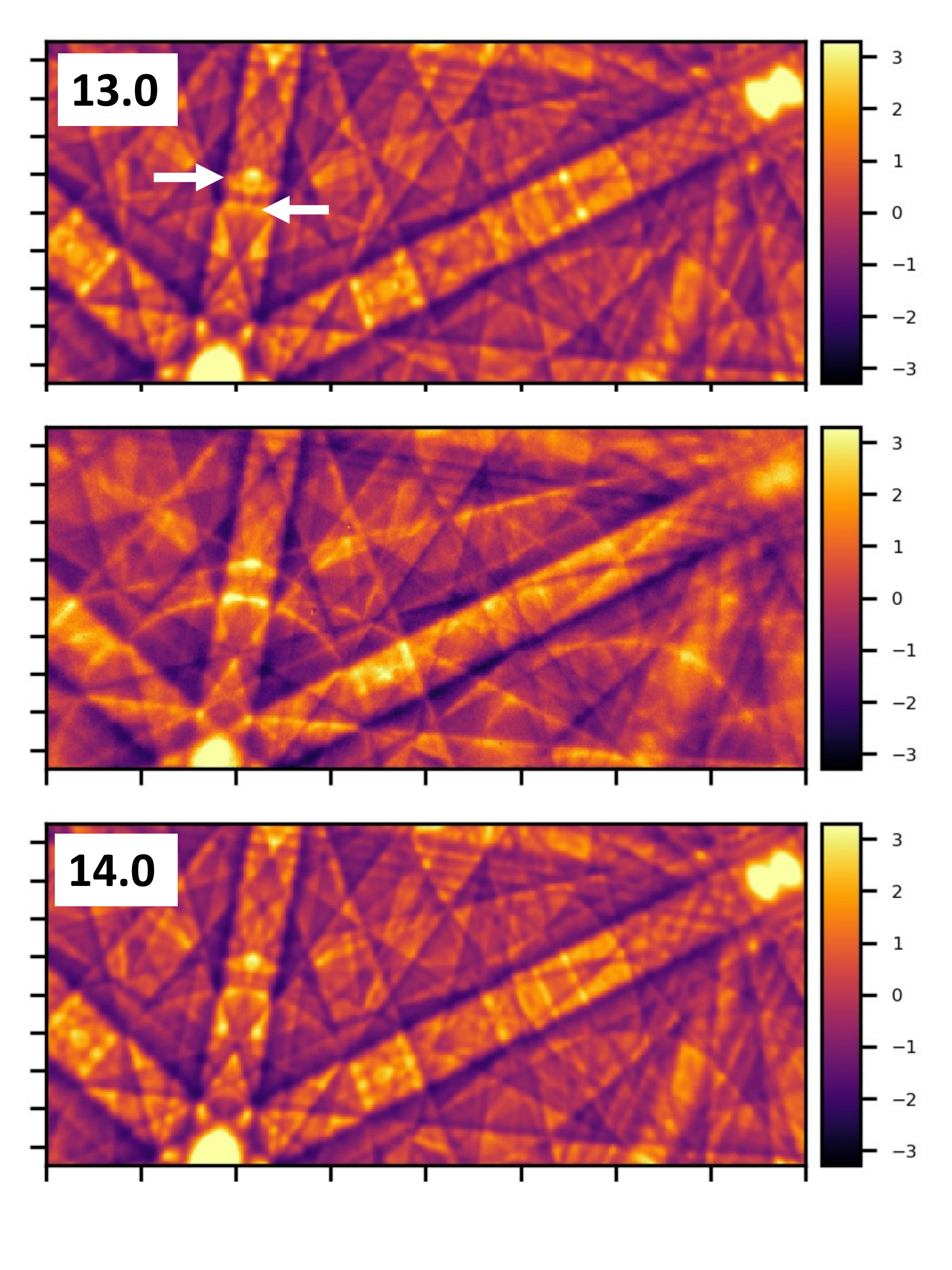}
    \caption{Analysis of the experimental middle ROI. Top: simulation for 13.0\,keV ($r=0.811$) with indicated features for comparison to the experiment. Center: Experimental middle ROI. Bottom: best fit ROI simulation for 14.\,keV central energy ($r=0.844$), features indicated in the upper panel fit better at 14\,keV.}
	\label{fig:roi_central}
\end{figure}

\begin{figure}[tb!] 
    \includegraphics[width=0.9\textwidth]{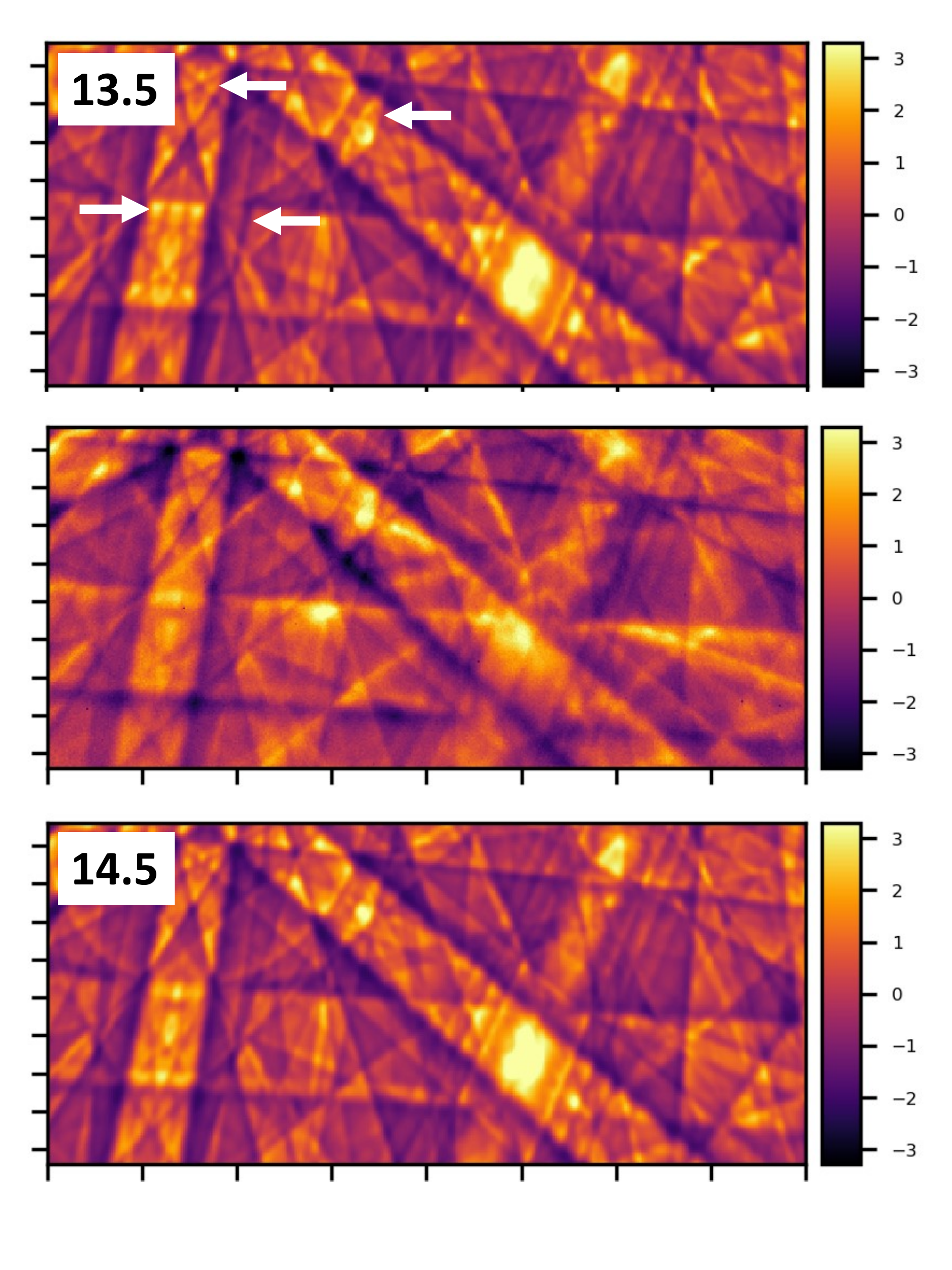}
    \caption{Analysis of the experimental lower ROI.
    Top: simulation for 13.5\,keV ($r=0.766$) with indicated features for comparison to the experiment. Center: Experimental middle ROI. Bottom: best fit ROI simulation for 14.5\,keV central energy (bottom, $r=0.792$), features indicated in the upper panel fit better at 14.5\,keV.}
	\label{fig:roi_bottom}
\end{figure}

We present a comparison of the energy-dependent pattern matching analysis for ROIs extracted from the experimental pattern (Fig.\,\ref{fig:sikiku}) in Figs.\,\ref{fig:roi_top}, \ref{fig:roi_central},  \ref{fig:roi_bottom}. These figures show matching at different energies (top and bottom panels, with the best fit energy in the bottom panel) against the experiment (middle panel). Features of interest are highlighted to guide the eye. These typically include high frequency features such as crossing band edges which are strongly dependent on the energy of the diffracting beams.

As can be seen in these Figures, a change by 1\,keV induces visible changes in the patterns, i.e. the more reasonable fit can be distinguished by eye within these limits.
This direct visual verification of the fit of specific features of crossing lines, as shown in Figures \ref{fig:roi_top}, \ref{fig:roi_central}, and \ref{fig:roi_bottom} is not influenced by slight changes in orientation or projection center, and the estimation of the mean effective energy should thus be very stable. 
Compared to the 1\,keV changes shown here, direct visual comparison by switching between patterns on the computer screen allows to distinguish the fits with about 500\,eV resolution (see supplementary data for comparisons).

\begin{figure}[tb!]  %
	\centering
	\includegraphics[width=10cm]{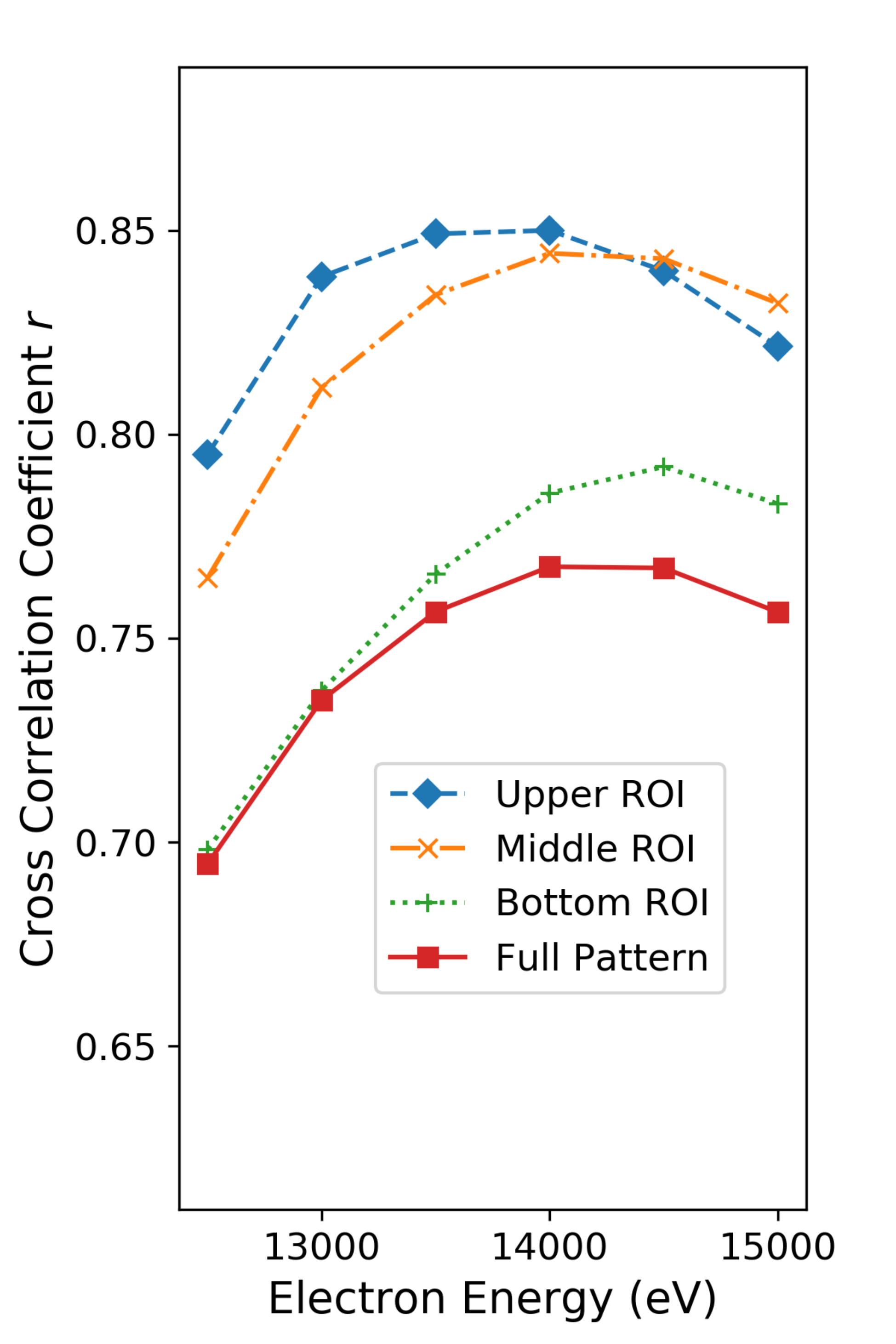}
	\caption{Energy-dependent NCC $r$-values for all ROIs and the full Si pattern as shown in Fig.\,\ref{fig:sikiku}. Lines are guides to the eye.}
	\label{fig:roi_fits}
\end{figure}

In Fig.\,\ref{fig:roi_fits}, we summarize the energy-dependent analysis of all the ROIs and the full Si pattern by carrying out an NCC optimization using energy-dependent master data for all energies between 12.5 and 15\,keV. 
We see that the NCC is peaked at energies between 13.5 and 14.5\,keV for all ROIs and the entire pattern match. There is a slight shift in the energy of the maximum NCC value depending on ROI position. The NCC is at the lowest energy for the upper ROI, which is has the largest scattering angle. The NCC is at the highest energy for the lower ROI, which has the lowest scattering angle. The NCC for the whole pattern falls between these two values.
We note that these peaked curves do not indicate a similar shape of the electron spectrum because changes in $r$ are not proportional to a corresponding change in spectrum. 
The NCC is very sensitive to changes in between patterns as it considers all features within the cross correlated region of interest, and even differences of $\Delta r = 0.01$ at values of $r>0.8$ indicate significantly worse fits. 
The lower ROI shows generally lower values of the NCC, which can be assigned to the strong excess-deficiency effects \cite{winkelmann2008um} which have not been included in the model, and this reduces similarity between the simulation and the experiment.

\begin{figure}[tb!]  %
	\centering
	\includegraphics[width=14cm]{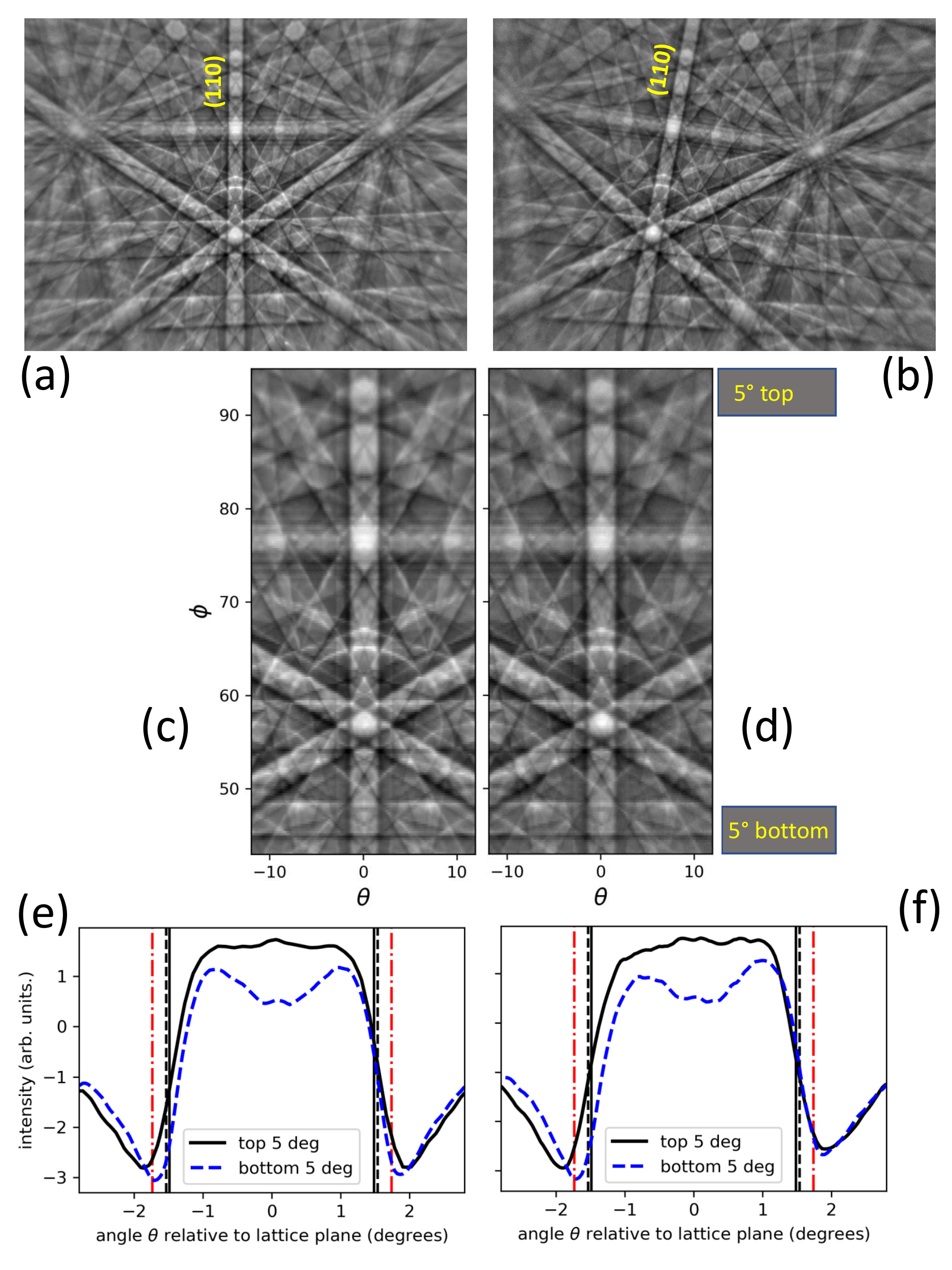}
	\vspace{-0.5cm}
	\caption{Comparison of (110) band profiles extracted from two Kikuchi patterns of slightly different orientations (a, b). The middle panels (c, d) show the (110) bands reprojected in a spherical coordinate system $(\theta,\phi)$ for quantitative band width comparison.
	Bottom (e, f): averaged (110) band profiles within 5\degree{} of the top and bottom of (c, d). The vertical lines indicate the geometrical Bragg angles for the 220 reflection in Si ($a_{Si}=5.4307$\,\AA, $d_{220}=1.920$\,\AA) for energies of 15\,keV (solid), 14\,keV (dashed), and 11\,keV (dash-dotted).}
	\label{fig:bands220}
\end{figure}

As an additional analysis option to further verify the results of the fitting procedure discussed above, we also extracted Kikuchi bands from the experimental patterns.
Because we know the gnomonic projection center calibration for the experimental pattern, we can reproject the experimental data to a spherical coordinate system $(\theta,\phi)$.
In this coordinate system, a Kikuchi band runs azimuthally (angle $\phi$) along the equator and extends to latitudes $\pm\theta$ as measured from the lattice plane trace at the equator with $\theta=0$. 
The extraction of a (110) band can be seen in Fig.\,\ref{fig:bands220}(c) and (d) for the gnomonic patterns shown in Fig.\,\ref{fig:bands220}(a) and (b), respectively. 
In the spherical coordinate system, we have access to the angles of Kikuchi band features relative to the lattice plane, i.e.  an angular broadening of the Kikuchi bands in Fig.\,\ref{fig:bands220}(c) and (d) can be directly detected. 

To this end, in Fig.\,\ref{fig:bands220}(e) and (f) we have plotted the averaged band profiles for the upper and lower 5\degree{} of the (110) bands as shown in the middle panels (c) and (d).
We indicated by vertical lines the geometrical Bragg angle for the 220 reflection in Si ($d_{220} = 1.920$\,\AA) for energies of 15\,keV (solid), 14\,keV (dashed), and 11\,keV (dash-dotted).
The Bragg angle is \textit{not} expected at the experimental minimum of the profile, but qualitatively in the vicinity of the inflection point of the shoulder, more like seen for 15\,keV and 14\,keV. This can be seen by a two-beam dynamical model of the Kikuchi band profile \cite{reimer1986sca}.
Because the dash-dotted vertical line for the 11\,keV Bragg angle is clearly outside the shoulder region, an energy change to 11\,keV is inconsistent with the current measurement for the solid profile from the top region of the pattern.

After a consistent pattern calibration and profile extraction, we have to fix the lattice plane position ($\theta = 0.0$), i.e. in general we are not free anymore to shift the $\theta$-profile in the plot. 
This is why it is significant to discuss the observed asymmetries which are related to the excess-deficiency effect caused by a mechanism that is related to the incident beam direction.
Negative angles correspond to the deficient edge of the profile, while the positive angles are on the excess side. We can see that the nearly vertical (110) band of the pattern in Fig.\,\ref{fig:bands220}(a) produces a much smaller asymmetry in the profiles than the inclined (110) band in Fig.\,\ref{fig:bands220}(b). 
These effects should be even stronger in the orientation which has been used in \cite{ram2018prb} where the (110) band is even more slanted, and within the work of Ram et al. the possible role of this effect has not been discussed. We note that the excess-deficiency effect is relatively small for the strong (110) bands in Si compared to other bands where this effect can be a significant part of the total experimental modulation. 
Irrespective of the possible influence of the excess-deficiency effect, however, we can see in the profiles of both patterns in the bottom panels, that an energy change to 11\,keV is inconsistent with the current measurement.

In summary, a direct band extraction and profile analysis implies that energies near 11\,keV and lower are incompatible with the measured profiles. 
The possible influence of the energy deficiency effect has to be considered for an experimental Kikuchi band profile analysis, especially if small changes in band widths are assumed to be relevant.   

In order to further investigate the role of the density of the material and the resolution of Kikuchi pattern features on the effective Kikuchi spectrum, we have carried a similar analysis for Kikuchi patterns measured from BaFe$_2$As$_2$ at a temperature of 12\,K and at a beam voltage of 20\,keV which contain a high density of high frequency diffraction information.

\begin{figure}[tb!]  %
	\centering
	\includegraphics[width=14cm]{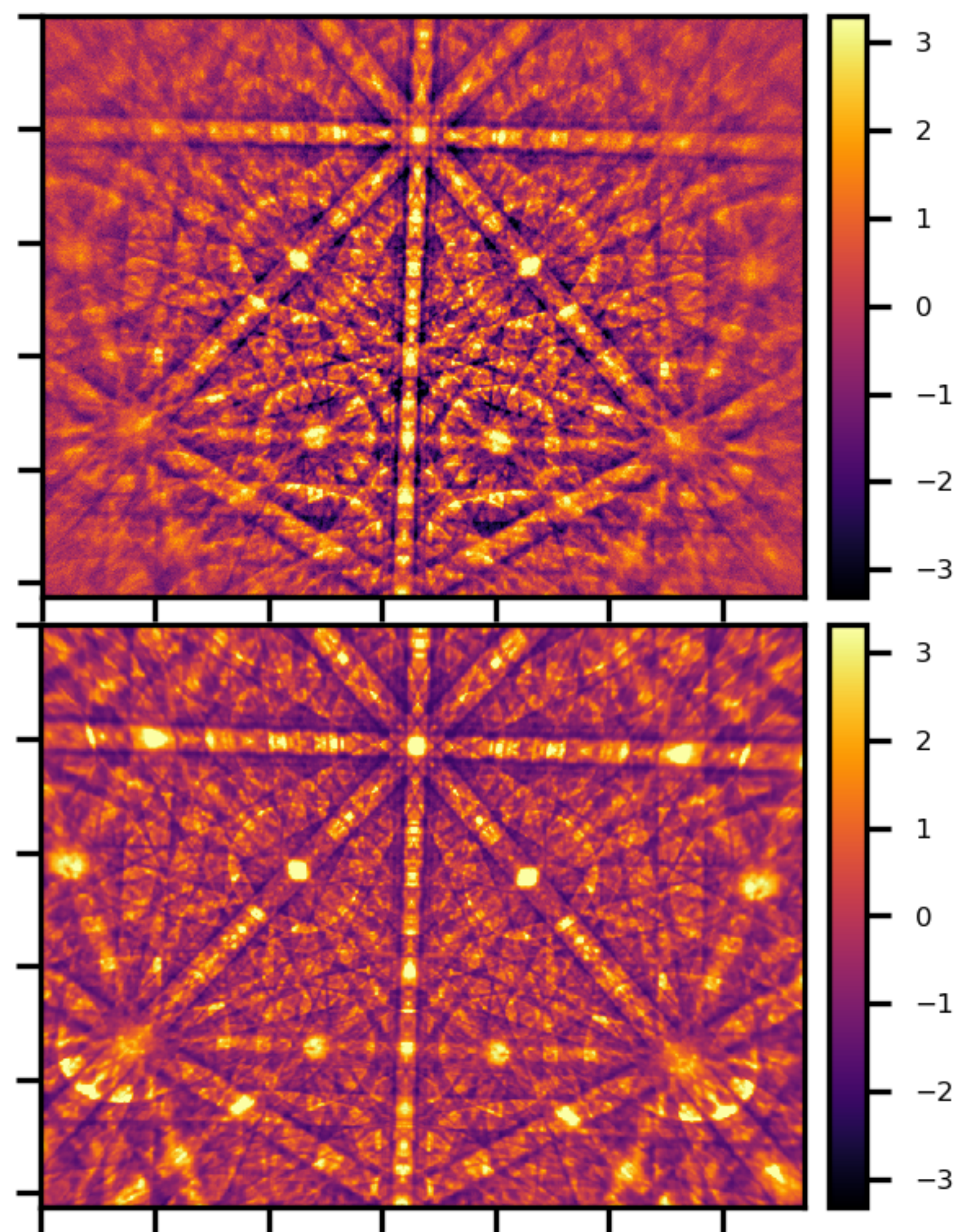}
	\caption{Top: Kikuchi pattern measured at 12\,K from BaFe$_2$As$_2$, primary beam voltage 20\,keV.\\
	Bottom: Best fit dynamical simulation for 20\,keV. The viewing angles are 89\degree{} horizontally, and 74\degree{} vertically.}
	\label{fig:bfarois}
\end{figure}

\begin{figure}[tb!]  %
	\centering
	\includegraphics[width=10cm]{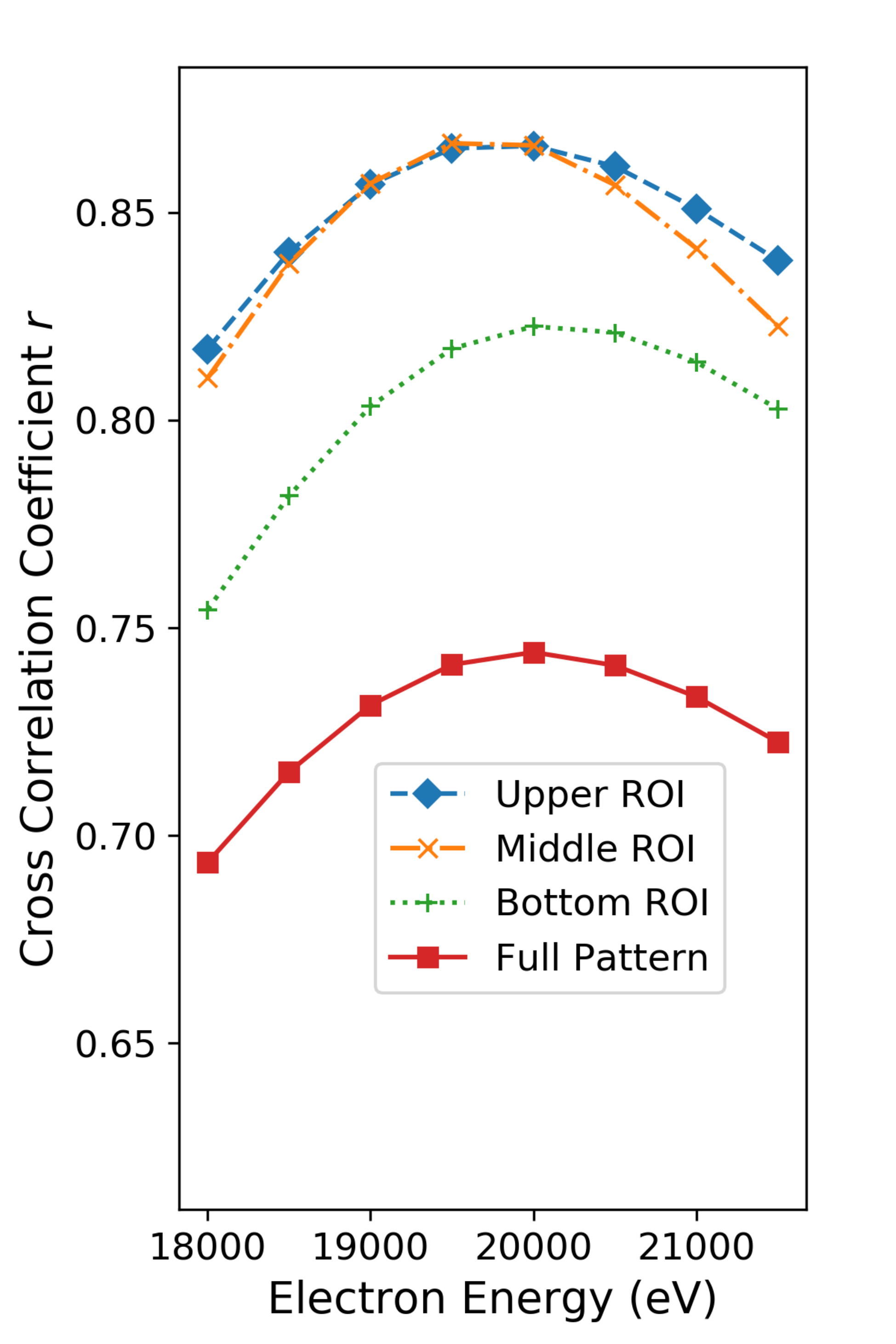}
	\caption{Energy-dependent NCC $r$-values for all ROIs and the full BaFe$_2$As$_2$ pattern as shown in Fig.\,\ref{fig:bfarois}. Lines are guides to the eye. }
	\label{fig:bfa_roi_fits}
\end{figure}

The results shown in Fig.\,\ref{fig:bfarois} and Fig.\,\ref{fig:bfa_roi_fits} show that for BaFe$_2$As$_2$ we obtain the optimum fit for all ROIs within approx. 500\,eV of the primary beam energy.
This trend is consistent with the dependence of the BSE spectra on the density and mean atomic number of a material, which is known to result in more intensity near the primary beam energy due to the increased probability of low-loss backscattered electrons via large-angle scattering events \cite{reimersem,goldsteinSEM4}.
The quantitative trend in the decrease of the shift of the mean effective energy relative the primary beam energy, as seen for silicon, GaN \cite{winkelmann2009ebsd2}, and BaFe$_2$As$_2$,  also seems to be inconsistent with the findings of \citeauthor{callahan2013mm} in  \cite{callahan2013mm}, where changes of the mean energy from 28.5\,keV to 23\,keV from the bottom to the top of the EBSD phosphor screen are derived by CSDA Monte Carlo simulations for Ni (Fig. 6a in \cite{callahan2013mm}).

\section{Discussion}

We have shown that the features observed in experimental Kikuchi diffraction patterns of silicon at 15\,keV are consistent with the assumptions of the simulation model discussed in \cite{winkelmann2007um,winkelmann2009ebsd2,winkelmann2016iop}, which assumes that a relatively narrow range of energies ($\lessapprox 1$\,keV) is sufficient to simulate Kikuchi pattern formation for EBSD applications.

We observe that these findings are inconsistent with the results shown by \citeauthor{ram2018prb}, who use the continuous slowing down approximation in their Monte Carlo simulations to calculate the energy spread of the electrons which form the Kikuchi bands. These simulations suggest much larger changes in the mean energy across and EBSD Kikuchi pattern, and also a much broader effective spectrum (several keV in both cases) than are consistent with our analysis.

In the following, we will discuss several issues which are related to (a) the use of an observed width of interference features to infer properties of the respective spectrum
(b) use of an inadequate simulation model for description of the energy spectrum.

\subsection{Experimental Factors affecting EBSD pattern resolution.}

In general, estimation of an electron energy spectrum from diffraction patterns can be highly unreliable, because the observation of broadened interference features does not necessarily imply a broadened spectrum.

Instead, we note that several factors can lead to a broadening of interference features in Kikuchi patterns when using a conventional setup like described in Fig.\,\ref{fig:s_ebsd}:

\begin{itemize}

  \item The detector response (modulation transfer function \cite{britton2013um,britton2010um}) due to the energy-dependent properties of the detector screen, the optical system, and the CCD/CMOS camera used for capturing the EBSD raw pattern can result in changes in the spatial frequency of features (blurring), also with respect to position of the detector and the relative illumination (e.g. due to vignetting). While this changes the relative quality of regions within the pattern, in the absence of significant optical distortions, the relative position of features will not be affected.
  
  \item Minimal orientation changes of the material in the measured sample area will lead to a superposition of patterns which are slightly rotated with respect to each other, thus leading to an effective broadening of diffraction features. This is reduced when a single crystal of unstrained material is explored (such as the Si(001) semiconductor wafer samples shown here).
  
  \item The sample surface quality can affect the pattern quality due to crystal deformations, defects, amorphous oxide layers and carbon contamination. 
  
  \item The quality of Kikuchi pattern features is influenced by temperature effects \cite{bastos2018mc} such as increased thermal vibrations.
  
  \item Diffraction of electrons from sources in thin ($<5$nm) regions of the surface will lead to a broadening of features due to reduced dynamical scattering effects \cite{winkelmann2010jm,winkelmann2010njp}. Decoherence of the electron beams in the Bloch waves will lead to reduction in diffraction modulation and an increase of the diffuse background signal. For emission from deeper parts of the crystal, anomalous absorption in the dynamical diffraction process can lead to a contribution with inverse contrast \cite{winkelmann2010um,brodu2017um} of the Kikuchi band profiles.
  The relative influence of these dynamical diffraction effects can change with scattering angle.
  
   \item The primary energy reference is not known exactly. This can be due to an uncertainty in how precisely the SEM voltage and the energy spread of the incoming electron beam is known, or by charging effects which change the actual landing energy on the sample.

\end{itemize}

From the results presented in Fig.\,\ref{fig:bands220}, we see that utilization of Kikuchi band widths to interpret the energy spread within a detected diffraction pattern can be problematic.
Even in two-beam dynamical electron diffraction theory the exact position of the geometrical Bragg angle is not given by a fixed physical feature in the band profile. For example, the position of the minimum of a Kikuchi band profile changes according to the structure factor of the relevant reflection and its absorption parameters even for a fixed energy \cite{reimer1986sca}. 
In experimental EBSD patterns, the interpretation of the band profile can be further complicated by band asymmetry due to the excess-deficiency effects and the due to the systematic distortion effects by the gnomonic projection. 
All these factors can make it difficult to precisely define a repeatable "width" of an experimental Kikuchi band, and we cannot be sure e.g. that changes in the position of a minimum somewhere in a Kikuchi band profile are related only to changes in energy.

Ultimately, the combination of items that impact the precise band width and resolution at different positions on the detector can be complicated. In the absence of a quantification of the various possible broadening mechanisms, extraction of Kikuchi band widths leads to inconclusive results (compare also 
Fig.\,3(a) and (b) in \cite{ram2018prb}, where all the main bands in the simulations can be seen to be slightly broader than in the experiment, but good agreement with experimental band widths extracted via the so-called "Kikuchi bandlet method" \cite{ram2014jac} is claimed nevertheless).

\subsection{Adequacy of Monte-Carlo simulations for EBSD applications}

Compared to previous dynamical simulation approaches for EBSD \cite{winkelmann2007um,villert2009jm}, the significant new feature of the simulation approach described in \cite{callahan2013mm,ram2018prb,singh2018srep} and applied by Ram et al.\@ is the quantitative use of a Monte Carlo simulation of electron trajectories to assign the full energy- and angle-dependent intensity of the electrons which are scattered towards the phosphor screen, i.e. the intensity in the top raw pattern of Fig.\,\ref{fig:sikiku}, comprising both the diffuse background and the additional Kikuchi diffraction modulation. 
To describe the inelastic scattering effects, the authors of \cite{callahan2013mm,ram2018prb,singh2018srep} use the continuous slowing down approximation (CSDA), in which the energy of an electron reduces continuously with the travelled path length, while the specific effects due to the actual discreteness of energy loss processes are disregarded.

It has been shown that the CSDA is a very good approximation for many applications in electron microscopy within a regime of large collision numbers \cite{werner2001sia,werner2010jesrp,dapor2017tees}.
In this ``slowing-down regime'', we can expect that the statistical fluctuations of discrete energy losses will average out and the energy loss increases continuously with the travelled path length, describing to a good approximation, for example, total backscattering yields, or the yields of excited X-rays \cite{joy1995mc}.
However, the CSDA fails substantially in electron spectroscopic applications at small collision numbers and small energy losses, the ``quasi-elastic regime'' \cite{shimizu1975jpd,shimizu1976jpd,ding1996scanning,werner2001sia,werner2010jesrp}. 
One of the most severe failures of the CSDA is the lack of an elastic peak in the electron spectrum, by which we can directly see that the CSDA cannot be appropriate for the analysis of experiments were the observed effects rely on the elastically scattered electrons and electrons with small energy losses. 
The regime of small collision numbers and small energy losses is especially relevant for diffraction effects, as we have seen from the experimental spectroscopic data \cite{winkelmann2010njp,vos2016um}. 
This can also be rationalized via theoretical arguments, because we can expect that electron waves involved in multiple scattering processes will become more and more incoherent with an increasing number of inelastic collisions in which their mutual phase relationships will be randomized.    
Because the CSDA is an approximation which fails at qualitatively reproducing the backscattered electron spectrum in the energy region that is actually highly relevant for Kikuchi diffraction effects, it is difficult to rely quantitatively on parameters extracted from simulations using this approximation.

Experimentally, the failure of CSDA-MC simulations for spectroscopic SEM investigations has been verified using an electrostatic energy filter with with 0.55\% energy resolution \cite{berger1999jesrp} to measure BSE spectra  for aluminum, silver, and gold for an incident beam angle of 80\degree{} and a range of scattering angles \cite{berger2000thesis}  (Fig.\,4.29, p.97), In \cite{berger2000thesis}, \citeauthor{berger2000thesis} found that a MC simulation based on continuous energy losses could not reproduce the shape of the measured BSE spectra, compared to simulations using statistical discrete energy losses, which provided good agreement (\cite{berger2000thesis} Fig.4.25, p.94).
In the analysis of their results using a high-pass energy filter, \citeauthor{deal2008um} applied CSDA-MC simulations and showed in Fig. 10 of \cite{deal2008um} that these CSDA simulations predict a significantly faster reduction of the accumulated BSE intensity at low energy losses than experimentally measured (implying that the CSDA underestimates the low-loss part of the spectrum). 

Also in investigations concerning the role of electron backscattering from silicon detectors in particle physics applications, it has been found that Monte Carlo simulations based on discrete inelastic processes are necessary to correctly describe the detector response \cite{renschler2011thesis,furse2017njp}.

Going beyond the CSDA by using discrete inelastic loss processes, it was shown in \cite{winkelmann2013mm} that MC simulations based on the differential inverse inelastic mean free path (DIIMFP) \cite{werner2001sia,werner2010jesrp,salvatpujol2013sia}, can reproduce a qualitatively correct energy spectrum including the elastic peak and distinct plasmon loss peaks. 
The trajectories of the electrons were analyzed in terms of the recoil energy in the scattering process in an attempt to obtain an estimation of the depth dependence of the Kikuchi diffraction source strength. The resulting depth distributions for the Kikuchi pattern electrons were of exponential decay type with decay constants of the order of the IMFP \cite{winkelmann2013mm}.
Also the angular distribution can be treated using this approach \cite{winkelmann2016iop}.

In the EBSD literature, Monte Carlo simulations have been applied by several authors for the analysis of EBSD experiments, see for example  \cite{ren1998mm,tao2004mm,deal2005sia,rice2017micron,gazder2017micron}. With respect to the discussion above it seems to be pertinent to state that the use of the CSDA for the quantitative simulation of diffraction effects and their interpretation in terms of electron trajectories should be considered as unreliable. 
In contrast, the conditions of the slowing-down regime are probably better fulfilled for a description of the diffuse background signal.
Although the diffuse background signal is usually removed from the experimental data, it contains a considerable amount of useful information about the sample \cite{winkelmann2017jm,britton2018mc}. 
Simulations of the diffuse background can also be helpful to estimate general trends in the necessary pattern collection times for a specific signal-to-background ratio in the presence of noise. 

Concerning the depth-dependent Kikuchi diffraction source intensity, it has been discussed in \cite{winkelmann2010jm} that the explanatory power of Monte Carlo simulations for the depth profiles of the Kikuchi pattern sources is limited by the physical mechanism of Kikuchi pattern formation itself: Due to the summation of individual diffraction patterns over an extended depth range, the resulting, measured, Kikuchi pattern shows a reduced sensitivity to the exact details of the depth profile and a considerable variation in the emission profiles can be compatible with an experimental pattern.
This is why we can obtain sufficiently good agreement between experimental and simulated patterns as discussed above by using rather general models for the depth distributions of the Kikuchi pattern electrons, i.e. by assuming parameterized profiles in the shape of exponential or Poisson distributions.

\section{Summary}
We used full pattern dynamical electron diffraction simulations to explore the possible impact of the backscattered electron energy spectrum on the appearance of EBSD patterns from silicon. 
We found that Kikuchi patterns from silicon are consistent with mean energies which are approximately 1 to 1.5\,keV below the primary beam energy, compared to a corresponding range between 2 and 5\,keV predicted in \cite{ram2018prb}. This is supported by an analysis of Kikuchi patterns from a crystalline material with higher mean atomic number, BaFe$_2$As$_2$. In both examples, we have evaluated not only the band width of one particular feature, but we have analyzed the appearance of extended and correlated Kikuchi pattern features, which provides greater precision in estimating the constraints on the effective Kikuchi pattern spectrum.
The experimentally observed broadening of diffraction features is consistent with narrow effective Kikuchi spectra (FWHM $\lessapprox 500$\,eV).
For higher-Z materials, the mean energy will approach the primary beam energy.
We find that the use of the continuous slowing down approximation (CSDA) in the Monte Carlo part of the Kikuchi pattern simulation approach presented in \cite{callahan2013mm,ram2018prb,singh2018srep} leads to inaccurate predictions for the electron energy spectrum that is effective in typical EBSD Kikuchi diffraction patterns. 
If we wish to consistently include the spatial origin and spectral properties of the Kikuchi pattern electrons, we suggest that Monte Carlo simulations based on discrete inelastic loss processes are explored according to the detailed electronic properties of the investigated materials, possibly in combination with consistent quantum-mechanical simulations of electron trajectories in the presence of diffraction \cite{cheng2018njp}.

\section*{Acknowledgements}
We thank P. Chekhonin, A. Pukenas, E. Hieckmann (Technical University Dresden, Germany) for providing the BaFe$_2$As$_2$ Kikuchi patterns. TBB thanks the Royal Academy of Engineering for financial support of his Research Fellowship. We thank Alex Foden for providing preliminary diffraction data.
We thank R. Saliwan-Neumann (BAM) for additional EBSD measurements.

\section*{Data Statement}
The data which has been generated in this study is available via a Zenodo/GitHub repository.


%

\end{document}